\documentclass[amsmath,amssymb,reprint,groupedaddress,showpacs,aps,pra,nofootinbib]{revtex4-1}
\usepackage{graphicx}
\usepackage{dcolumn}
\usepackage{bm}
\usepackage{color}
\usepackage{outlines}
\usepackage{amsmath} 
\usepackage{gensymb}
\definecolor{red}{rgb}{1,0.,0}

\usepackage{hyperref}
\newcommand{\ket}[1]{\left| #1 \right>} 

\begin{document}

\title{A tamper-indicating quantum seal}

\author{Brian~P.~Williams}
\affiliation{Quantum Computing Institute, Oak Ridge National Laboratory, Oak Ridge, Tennessee USA 37831}
\email{williamsbp@ornl.gov}
\author{Keith~A.~Britt}
\affiliation{Quantum Computing Institute, Oak Ridge National Laboratory, Oak Ridge, Tennessee USA 37831}
\email{brittka@ornl.gov}

\author{Travis~S.~Humble}
\affiliation{Quantum Computing Institute, Oak Ridge National Laboratory, Oak Ridge, Tennessee USA 37831}
\email{humblets@ornl.gov}

\begin{abstract}
Technical means for identifying when tampering occurs is a critical part of many containment and surveillance technologies. Conventional fiber optic seals provide methods for monitoring enclosed inventories, but they are vulnerable to spoofing attacks based on classical physics. We address these vulnerabilities with the development of a quantum seal that offers the ability to detect the intercept-resend attack using quantum integrity verification. Our approach represents an application of entanglement to provide guarantees in the authenticity of the seal state by verifying it was transmitted coherently. We implement these ideas using polarization-entangled photon pairs that are verified after passing through a fiber-optic channel testbed. Using binary detection theory, we find the probability of detecting inauthentic signals is greater than 0.9999 with a false alarm chance of $10^{-9}$ for a 10 second sampling interval. In addition, we show how the Hong-Ou-Mandel effect concurrently provides a tight bound on redirection attack, in which tampering modifies the shape of the seal. Our measurements limit the tolerable path length change to sub-millimeter disturbances. These tamper-indicating features of the quantum seal offer unprecedented security for unattended monitoring systems.
\end{abstract}
\pacs{42.50.Ar,03.67.Bg,42.79.Sz}
\maketitle
\section{Introduction}\label{introsec}
Tamper-indicating optical seals are widely used for verifying the integrity of enclosed systems, including storage containers, physical perimeters, and fiber networks \cite{Griffiths1995,Chtcherbakov1998,Szustakowski2005,Juarez2005}. Fiber-optic seals have proven especially useful for actively surveying large areas or inventories due to the extended transmission range and flexible layout of fiber \cite{Horton1995,Johnston2001}. These seals operate as optical continuity sensors that confirm transmission of an encoded light pulse from source to receiver with tampering indicated by either the absence of the light or an error in the received encoding.
\par
In the classical setting, detection of tampering requires failure of the intruder to accurately replicate the original transmission. This is typically accomplished with ``secret'' information that is hidden from the intruder, for example, the optical modulation sequence used to transmit pulses. This secret information, however, is vulnerable to discovery by the intruder using conventional signal detection methods. Thus, in principle, an attacker is  able to perfectly replicate the optical signal using {\it a priori} knowledge, and the classical variant of an optical seal is vulnerable to an intercept-resend spoofing attack. In this case, the intruder has the ability to recover information, such as the frequency, bandwidth, and modulation, that describes the classical state of the light. An exact duplicate of the transmitted signal can then be replicated by the attacker and injected into the fiber, thus spoofing the downstream sensor.
\begin{figure}[b]
\begin{center}
\includegraphics[scale=0.37]{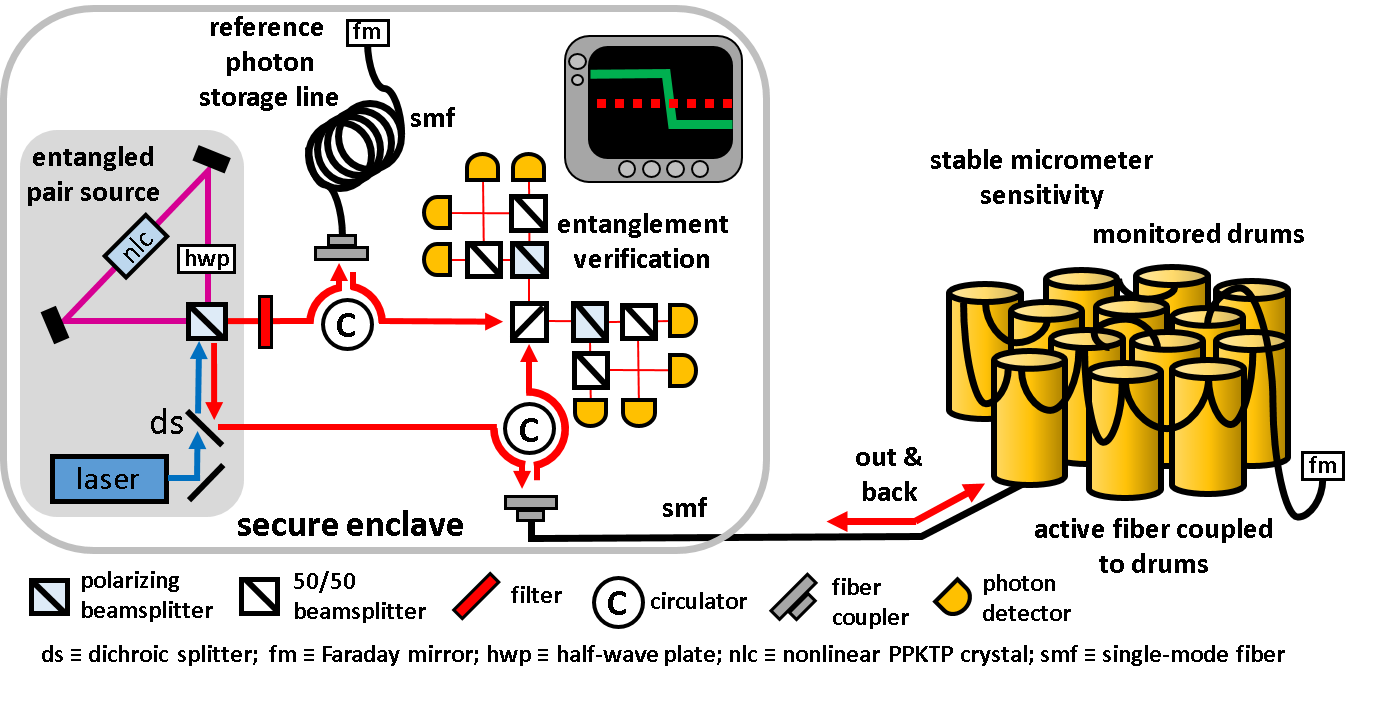}
\caption{A schematic of a tamper-indicating quantum optical seal showing four major components: (a) an entangled photon source, (b) reference and active fiber optic links, (c) an entanglement verification measurement, and (d) a monitoring system that process timestamped single photon detections.}
\label{figschematic}
\end{center}
\end{figure}
\par
By contrast, cloning quantum information is prohibited by the linearity of quantum mechanics \cite{Wootters1982,Dieks1982}. Attempts to clone quantum information, even optimally, necessarily introduce noise into the resulting state and its subsequent observables \cite{Scarani2005}. These guarantees of the no-cloning theorem are well known from quantum key distribution (QKD), where non-local correlations inherent to quantum states are used to secure correlated measurements outcomes between users \cite{Gisin2002}. A QKD eavesdropper's attempt to clone transmitted quantum states introduces additional noise into the observed results that reveal her presence to the users \cite{Lutkenhaus1996,ShorPreskill2000,Bartkiewicz2013}. In the context of QKD, added noise manifests as larger bit-error rates in the raw measurements and theoretical considerations have set upper limits on the security of these systems.
\par
We report how the no-cloning theorem can be applied to the development of tamper-indicating seals. In particular, we close the intercept-resent vulnerability for optical seals by using entangled quantum states to monitor signal continuity, and we demonstrate how the intercept-resent attack can be detected by monitoring the entanglement between a pair of transmitted photons. Because entanglement between the photon pair can only be generated at the authenticated transmitter, any attempt to spoof the receiver with a replicated photon is immediately detected. We formalize detection of tampering in terms of a statistical estimate of the entanglement and show how entanglement distinguishes between an authentic seal state and a potential tampered seal. We implement these ideas using spectrally and polarization-entangled photon pairs as part of a quantum signal detection system. Our results find a probability of detection for inauthentic signals greater than 0.99999 at a false alarm rate of one in a billion for a 10 second sampling interval.
\par
We also present tight bounds on an intruder's ability to spoof the seal using a second type of attack based on redirection of the signal path. The latter vulnerability arises when the intruder changes the path length of the optical signal, perhaps when rerouting around the intended enclosure. Conventional approaches to detect this attack rely on anomalies in the signal time-of-arrival, which is limited by the timestamping electronics. We employ the Hong-Ou-Mandel effect between energy-degenerate entangled photons to detect submillimeter path length changes. Finally, we present a prototype implementation and  experimental results that validate detection of these various tampering methods with estimations of the transmitted entanglement.
\par
Entanglement has been proposed previously to offer an advantage for various forms of tamper detection. This includes our earlier efforts to demonstrate intrusion detection by monitoring the free-space optical transmission of entangled states as a kind of a quantum tripwire \cite{Humble2009}. Those experiments used a sensing configuration in which the polarization entanglement between two photons transmitted to different measurement receivers was quantified. This system was subsequently improved with the development of non-local polarization interferometry, which offers greater sensitivity to the transmitted state while providing more efficient sampling  \cite{Williams2014}. Similar research for perimeter monitoring has applied the idea of interaction-free measurement to create an invisible quantum fence, in which the intruder is unable to detect the quantum signal state with very high probability \cite{anisimov2010invisible}. There have also been previous uses of entanglement verification applied to remote sensing, where the quantum statistics of a reflected signal are used to authenticate the results of an imaging system \cite{malik2012quantum}.
\par
The paper is organized as follows. In Sec. \ref{theorysec}, we present our model for the quantum  seal, and we formalize the problem of tamper detection using binary detection theory. In Sec.~\ref{experimentsec}, we detail the experimental implementation of the seal including the entangled photon pair source, entanglement detection, and Hong-Ou-Mandel interference measurements. Additionally, the results of our experimental studies to characterize seal performance in term of observed entanglement statistics and acquisition times are given in this section. We conclude our presentation in Sec.~\ref{conclusionsec}.
\section{Theoretical Model}\label{theorysec}
A schematic diagram of a quantum seal prototype is presented in Fig.\ref{figschematic}. It consists of a secure enclave that generates and detects a quantum signal state and an unsecure area that corresponds with the system to be monitored. The quantum signal state is a pair of photons prepared in a polarization-entangled Bell state by cw-pumping of type-{II} spontaneous parametric down-conversion (SPDC). These photons are coupled into active and reference fibers, where the active fiber traverses the unsecure area and the reference fiber remains inside the secure enclave. 
\par
\begin{figure}[tb]
\centering
\includegraphics[scale=0.3]{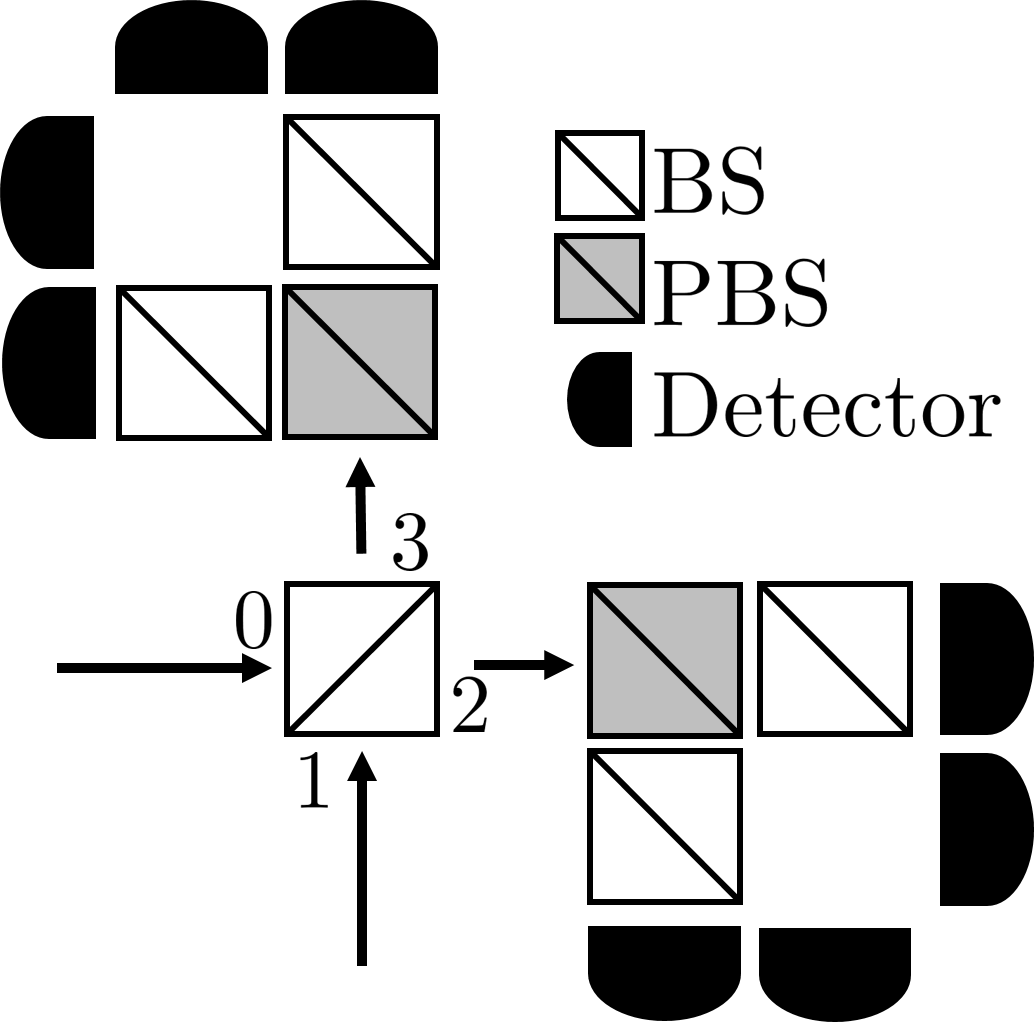}
\caption{A Bell-state analyzer (BSA) consists of two modes labeled as $0$ and $1$ input to a symmetric beamsplitter, whose output modes $2$ and $3$ direct to a pair of polarization analyzers and single-photon detectors. A beamsplitter after the polarization analyzer allow probabilistic detection of each coincidence type.
\label{BSA}}
\end{figure}

The footprint of the active fiber is configured to monitor access to an enclosure, for example, the inventory of closed containers shown in Fig.~\ref{figschematic}. The reference fiber is co-located with the transmitter and effectively represents a photon storage loop. Both fibers terminate at a Faraday mirror that reflects the photons and corrects for any polarization scrambling prior to the return tip. 
After reflection, each photon is router by a circulators to one of the input ports of the Bell-state analyzer (BSA) shown in Fig.~\ref{BSA}.  The photons are subsequently detected and the resulting information output port and polarization enables partial discrimination of the Bell states and entanglement detection.
\subsection{Entanglement Detection}
The Bell-state analyzer (BSA) shown in Fig.~\ref{BSA} supports partial discrimination between the polarization-encoded Bell states \cite{weinfurter1994,BraunsteinMann95},
\begin{align}
\left|\Psi^\pm\right\rangle&=(\left|H_0 V_1\right\rangle\pm\left|V_0 H_1\right\rangle)/\sqrt{2}\nonumber\\
 \left|\Phi^\pm\right\rangle&=(\left|H_0 H_1\right\rangle\pm\left|V_0 V_1\right\rangle)/\sqrt{2}\textrm{.}\nonumber
\end{align}
where $H_i$ and $V_i$ label horizontally and vertically polarized photons in spatial mode $i$. The states $\Psi^+$ and $\Psi^-$ have distinct detection signatures that allow deterministic identification, whereas the $\Phi^\pm$ states are not distinguishable amongst themselves but are distinguishable from $\Psi^\pm$ \footnote{A complete BSA is possible using linear optics and hyperentanglement \cite{Schuck2006complete,barbieri2007complete,barreiro2008beating}.} Moreover, consider an arbitrary two-photon pure state
\begin{align}
\ket{\theta}&=(a\ket{H_0, H_1} + b \ket{H_0, V_1} + c \ket{V_0, H_1} + d \ket{V_0, V_1} \nonumber\\
&=(a\hat{h}^{\dagger}_0\hat{h}^{\dagger}_1+b\hat{h}^{\dagger}_0\hat{v}^{\dagger}_1+c\hat{v}^{\dagger}_0\hat{h}^{\dagger}_1+d\hat{v}^{\dagger}_0\hat{v}^{\dagger}_1)\left|0\right\rangle
\label{state}
\end{align}
where the photons are temporally indistinguishable, operator $\hat{q}^{\dagger}_i\hat{r}^{\dagger}_s$ creates $q,r\in\left\{h,v\right\}$ polarization photons in spatial modes $i,s\in\{0,1,2,3\}$, $(1+\delta_{qr}\delta_{is})^{\textrm{-}1\!/2}\hat{q}^{\dagger}_i\hat{r}^{\dagger}_s\left|0\right\rangle=\left|Q_iR_s\right\rangle$, and $[\hat{q}_i,\hat{r}^{\dagger}_s]=\delta_{qr}\delta_{is}$ holds.
We calculate the expected detection signature for this state in the BSA by defining the operator transforms under the symmetric beamsplitter as \cite{Schuck2006complete,ou2007multi}
\begin{align}
\hat{h}^{\dagger}_0 \overset{BS}{\rightarrow}\left(\hat{h}^{\dagger}_2+i\hat{h}^{\dagger}_3\right)/\sqrt{2}&\quad\quad\hat{h}^{\dagger}_1 \overset{BS}{\rightarrow}\left(i\hat{h}^{\dagger}_2+\hat{h}^{\dagger}_3\right)/\sqrt{2}\label{bs1}\\
\hat{v}^{\dagger}_0 \overset{BS}{\rightarrow}\left(\hat{v}^{\dagger}_2-i\hat{v}^{\dagger}_3\right)/\sqrt{2}&\quad\quad\hat{v}^{\dagger}_1 \overset{BS}{\rightarrow}\left(-i\hat{v}^{\dagger}_2+\hat{v}^{\dagger}_3\right)/\sqrt{2}
\label{bs2}
\end{align}
where $h_0^{\dagger}$ is the creation operator for $\ket{H_0}$, etc. The corresponding post-beamsplitter state is
\begin{align}
\left|\theta'\right\rangle&\!=(1/2)[ia(\hat{h}^{\dagger 2}_2+\hat{h}^{\dagger 2}_3)-id(\hat{v}^{\dagger 2}_2+\hat{v}^{\dagger2}_3)+i(b-c)\hat{h}^{\dagger}_2\hat{v}^{\dagger}_2\nonumber \\
&-i(b-c)\hat{h}^{\dagger}_3\hat{v}^{\dagger}_3+(b+c)\hat{h}^{\dagger}_2\hat{v}^{\dagger}_3+(b+c)\hat{v}^{\dagger}_2\hat{h}^{\dagger}_3]\left|0\right\rangle
\label{poststate}
\end{align}
and the complete set of detection probabilities
\begin{align}
P_{h_2h_2}&=P_{h_3h_3}=|a|^2/2\label{prob0}\\
P_{v_2v_2}&=P_{v_3v_3}=|d|^2/2\label{prob1}\\
P_{h_2v_3}&=P_{h_3v_2}=\left[|b|^2+|c|^2+(b^*c+bc^*)\right]/4\label{prob2}\\
P_{h_2v_2}&=P_{h_3v_3}=\left[|b|^2+|c|^2-(b^*c+bc^*)\right]/4\textrm{.}
\label{prob3}
\end{align}
As required, the sum over all probabilities is unity. Coincidence probabilities involving different polarizations are related to the overlap of the input state in Eq.~(\ref{state}) with $\Psi^{\pm}$, i.e.,
\begin{align}
P_{h_2v_3}+P_{h_3v_2}&= |\left\langle\theta|\Psi^+\right\rangle|^2\\
P_{h_2v_2}+P_{h_3v_3}&=|\left\langle\theta|\Psi^-\right\rangle|^2\textrm{.}
\end{align}

We use the observed BSA detection statistics to distinguish the Bell-state prepared by the secure enclave from any other possible signal state. We define the polarization-correlation parameter
\begin{equation}
\mathcal{E}\equiv P_{h_2 v_3}+P_{h_3 v_2}-P_{h_2 v_2}-P_{h_3 v_3}\label{correlation}
\end{equation}
which is positive (negative) when orthogonal polarizations at different (same) ports is most probable. For the monochromatic pure state of Eq. (\ref{state})
\begin{equation}
\mathcal{E}=b^*c+bc^*\textrm{.}\label{monochromatic}
\end{equation}
The utility of $\mathcal{E}$ is that it places tight lower bounds on the strength of the correlations expected from the received states. For example, consider the separable state
\begin{equation}
\left|\theta_s\right\rangle\!=\!(\cos\alpha\hat{h}^{\dagger}_0+e^{iA}\sin\alpha\hat{v}^{\dagger}_0)(\cos\beta\hat{h}^{\dagger}_1+e^{iB}\sin\beta\hat{v}^{\dagger}_1)\left|0\right\rangle
\end{equation}
for which
\begin{align}
a=\cos\alpha\cos\beta \quad&\quad b=e^{iB}\cos\alpha\sin\beta \\
c=e^{iA}\sin\alpha\cos\beta\quad&\quad d=e^{i(A+B)}\sin\alpha\sin\beta\textrm{,}
\end{align}
in Eq.~(\ref{state}) with phases $\alpha, \beta, A, B \in [0,2 \pi]$. This restricted state leads to the correlation parameter
\begin{equation}
\mathcal{E}_s=\sin2\alpha\sin2\beta\cos(A-B)/2
\end{equation}
that is strictly bounded as
\begin{equation}
-1/2\leq\mathcal{E}_s\leq 1/2\textrm{.}\label{bound}
\end{equation}
More generally, any separable mixture
\begin{equation}\rho_s=\sum_k w_k \left|\theta^k_s\right\rangle\left\langle\theta^k_s\right|\textrm{,}\nonumber\end{equation}
with $w_k \geq 0$ and $\ket{\theta_k}$ the $k$-th pure state
also satisfies this bound, since
\begin{equation}
\left|\sum_k w_k \mathcal{E}^k_s\right|\leq \sum_k w_k/2 = 1/2\textrm{.}
\end{equation}
By comparison, the entangled states $\Psi^+$ and $\Psi^-$ have $\mathcal{E}=1$ and $\mathcal{E}=-1$, respectively, while the cross-correlation parameter vanishes for the $\Phi^{(\pm)}$ states. Thus, $\mathcal{E}>1/2$ indicates an entangled state, but $\mathcal{E}$ is not an entanglement metric. The sharp distinction in the parameter $\mathcal{E}$ permits us to classify an arbitrary input state as either $\Psi$-like entangled or not. 
\par
The analysis above neglects the finite temporal duration of each single-photon wavepacket in favor of a simplified monochromatic representation of the polarized states. In order to accurately model the time-of-arrival for each photon, we represent the Bell-state generated by the source as a multi-mode entangled photon pair
\begin{align}
\left|\Psi^+\right\rangle\!\!=&\frac{1}{\sqrt{2}}\!\!\!\int\! d\omega \!\!\int \!\!  d\omega' f(\omega,\omega') \nonumber \\
&\times \left[\hat{h}_0^{\dagger}(\omega)\hat{v}_1^{\dagger}(\omega')+\hat{v}_0^{\dagger}(\omega)\hat{h}_1^{\dagger}(\omega')\right]\left|0\right\rangle\textrm{.}
\end{align}
where $f(\omega,\omega')$ is the joint spectral amplitude of the photon pair and $\hat{j}_i^{\dagger}(\omega)$ creates a single $j\in\left\{h,v\right\}$ polarization photon in spatial mode $i\in\{0,1,2,3\}$ with frequency $\omega$. For these continuous operators the commutation relation is
\begin{equation}
[\hat{j}_i(\omega)\textrm{,}\hat{\ell}^{\dagger}_s(\omega')]=\delta_{j\ell}\delta_{is}\delta(\omega-\omega')
\end{equation}
The symmetric beam splitter relations are the same as in the single frequency case, Eq.~(\ref{bs1}) and (\ref{bs2}), and instead of the state given in Eq.~(\ref{poststate}), we have the post-beamsplitter state
\begin{align}
\left|\theta_{\Psi^+}\right\rangle=\frac{1}{2\sqrt{2}} \int\!d\omega\!\!\int & d\omega' f(\omega,\omega')e^{i\omega't_d}\times \nonumber \\
&\!\!\!\!\left(-i[\hat{h}^{\dagger}_2(\omega)\hat{v}^{\dagger}_2(\omega')-\hat{v}^{\dagger}_2(\omega)\hat{h}^{\dagger}_2(\omega')]\right.\nonumber \\
&\!\!-i[\hat{v}^{\dagger}_3(\omega)\hat{h}^{\dagger}_3(\omega')-\hat{h}^{\dagger}_3(\omega)\hat{v}^{\dagger}_3(\omega')]\nonumber \\
&+[\hat{h}^{\dagger}_2(\omega)\hat{v}^{\dagger}_3(\omega')+\hat{h}^{\dagger}_2(\omega')\hat{v}^{\dagger}_3(\omega)]\nonumber \\
&\left.+[\hat{v}^{\dagger}_2(\omega')\hat{h}^{\dagger}_3(\omega)+\hat{v}^{\dagger}_2(\omega)\hat{h}^{\dagger}_3(\omega')]\right)\left|0\right\rangle\nonumber 
\end{align}
where $t_d$ is the potential delay between the reference and active photon arrival times. This delay arises when the path length of active photon is shortened $(t_d < 0)$ or lengthened ($t_d > 0)$. The multi-mode detection probabilities are now given as
\begin{equation}
P_{j_i \ell_s}=\int_T d\tau|\left\langle0\right|\hat{j}_i(t)\hat{\ell}_s(t+\tau)|\Psi^{+'}\rangle|^2
\end{equation}
where
\begin{equation}
\hat{j_i(t)}=\frac{1}{\sqrt{2\pi}}\int d\omega \hat{j}_i(\omega)e^{-i\omega t}\textrm{.}
\end{equation}

\begin{figure}[tb]
\begin{center}
\includegraphics[scale=0.7]{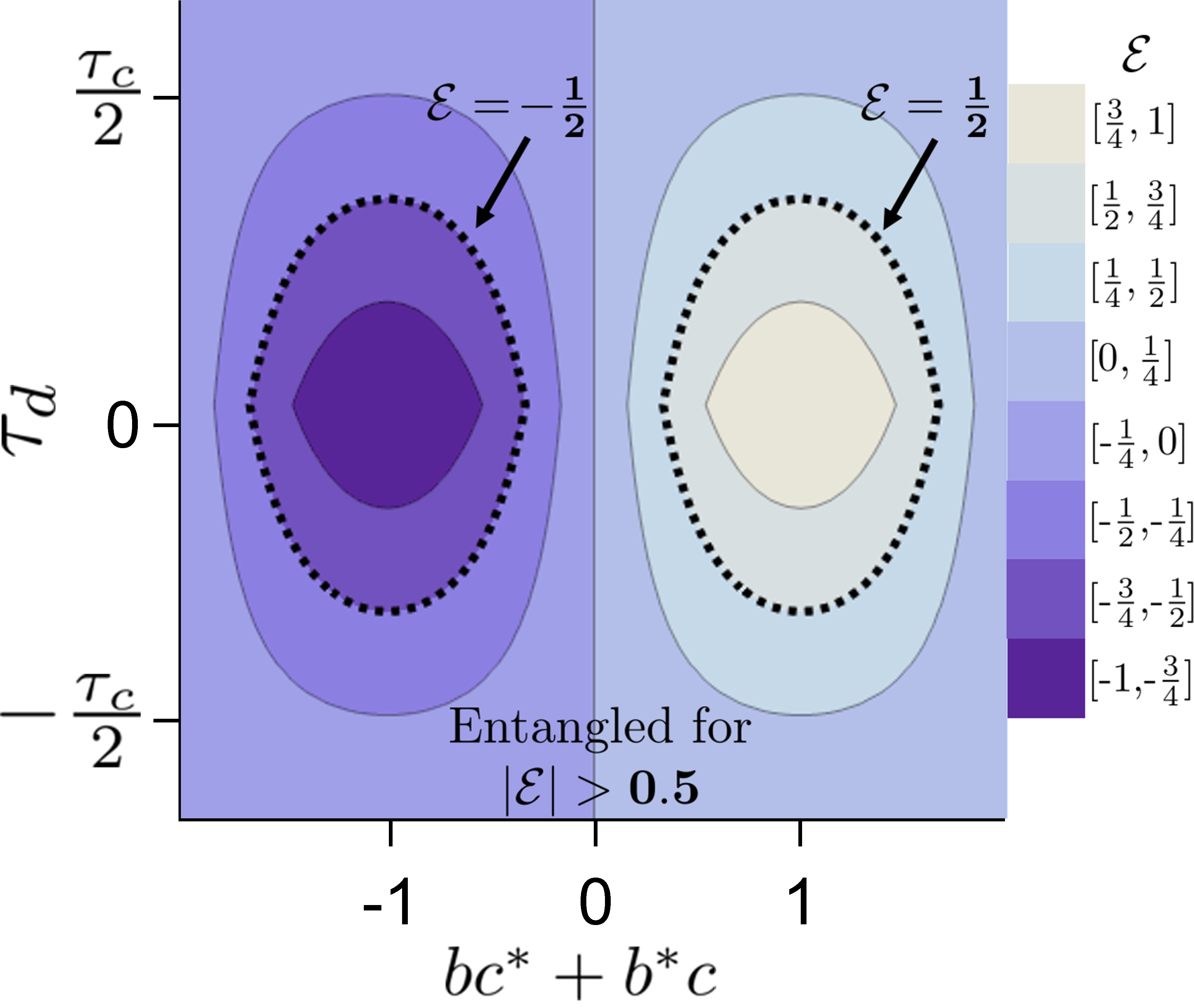}
\caption{A contour plot of the entanglement parameter $\mathcal{E}$ presented in Eq.~(\ref{correlation3}) with respect to the pure state coefficients $bc^*$+$b^*c$ and time delay $t_d$. The parameter $\mathcal{E}$ has extremal values when the input state is maximally polarized and there is no relative delay between the photons.}
\label{contour}
\end{center}
\end{figure}

The joint spectral amplitude for a cw-pumped type-II SPDC source is modeled as
\begin{equation}
f(\omega,\omega')=\delta(\omega+\omega'-\omega_p)\textrm{sinc}\left(\Delta k L/2\right)\end{equation}
where the longitudinal wave vector mismatch $\Delta k$
depends on the group velocity for the ordinary and extraordinary optical axes \cite{sergienko1995experimental,Grice1997}. In this regime, the detection probabilities become
\begin{align}
P_{h_2v_2}=P_{h_3v_3}&=\frac{1}{4}\left[1-{\textrm{\Large$\wedge$}}\left(2t_d/\Delta t\right)\right] \\
P_{h_2v_3}=P_{v_2h_3}&=\frac{1}{4}\left[1+{\textrm{\Large$\wedge$}}\left(2t_d/\Delta t\right)\right]
\end{align}
where $\Delta t$ is the propagation delay between ordinary and extraordinary photons travelling the full length of the nonlinear crystal \cite{sergienko1995experimental}.
\begin{equation}
\textrm{\Large$\wedge$}\left(x\right)=
\begin{cases}
   1-|x| & \text{if } |x| \leq 1 \\
   0 & \text{otherwise.}
\end{cases}
\end{equation}
These results converge to the monochromatic case when the delay $t_d = 0$, whereas if tampering delays the photon in the active fiber, but preserves the polarization entanglement, then the delay adds temporal distinguishability to the photons. For the multi-mode case with maximal polarization entanglement and temporal distinguishability, the cross-correlation parameter behaves as
\begin{equation}
\mathcal{E}_{pe}={\textrm{\Large$\wedge$}}\left(2t_d/\Delta t\right)\textrm{.}
\label{correlation2}
\end{equation}
Therefore, delays greater than a fraction of $\Delta t$ are detected even if the polarization entanglement is preserved. For our optical implementation, $\Delta t \approx 8$ picoseconds and a one-way path length difference greater than $300$ microns results in $|\mathcal{E}|\leq 1/2$. As one might expect, it can be shown that an arbitrary polarization multi-mode pure state has a correlation dependent on both the degree of polarization entanglement and temporal distinguishability,
\begin{equation}
\mathcal{E}_{mm}=
{\textrm{\Large$\wedge$}}(2t_d/\Delta t)(bc^*+b^*c)
\label{correlation3}
\textrm{.}
\end{equation}
For $t_d/\Delta t<<1$, we recover the monochromatic result presented in Eq.~(\ref{monochromatic}).
The entanglement dependency removes the possibility that the state may be spoofed, while the temporal sensitivity limits the adversary's ability to attack by redirection. In Fig.~\ref{contour}, the temporal and phase space of the parameter $\mathcal{E}$ is represented graphically.

\subsection{Entanglement Parameter Estimation}
We detect entanglement by estimating $\mathcal{E}$ from the correlated polarizations measured by the BSA. Due to the symmetry of the BSA, the coincident probabilities in Eqs.~(\ref{prob0})-(\ref{prob3}) satisfy
\begin{align}
P_{h_2 h_3}=&P_{v_2 v_3} = p_{sd}/2 \\
P_{h_2 h_2}=&P_{h_3 h_3}=P_{v_2 v_2}=P_{v_3 v_3} = p_{ss}/4\label{countreduce0}\\
P_{h_2 v_2}=&P_{h_3 v_3} = p_{ds}/2 \\
P_{h_2 v_3}=&P_{v_2 h_3} = p_{dd}/2
\end{align}
where subscripts $s$ and $d$ denote detections as being same or different, respectively, in the polarization and output mode. We estimate $\mathcal{E}$ from these four parameters.
\par
We estimate the set of coincident probabilities $\mathcal{P}=\left\{p_{sd},p_{ss},p_{ds},p_{dd}\right\}$ from the experimentally observed coincidence counts $c_{ij} \forall  i,j \in \{s,d\}$. However, the relative frequency of these measured rates do not directly correspond to the probabilities due to several technical limits. First, the SPDC source described in Sec. \ref{experimentsec} is known to result in registering accidental coincidences between photons from different entangled pairs and to a lesser degree photon-dark count coincidences. These so-called ``accidentals" are rare but add a contribution $c_{acc}$ to the observed coincidences. Second, the efficiencies for each joint two-photon pathway are typically different. Our first correction removes contributions from accidental coincidences and eliminates joint efficiency asymmetries with the formula
\begin{equation}
c_{i}'=\frac{\eta_{min}}{\eta_{i}}(c_{i}-c_{acc})\label{coincCounts}
\end{equation}
where $i$$\in$$\{$$h_2h_3$, $v_2v_3$, $h_2h_2$, $v_2v_2$, $h_2v_2$, $h_3v_3$, $h_2v_3$, $v_2h_3$$\}$, $c_i$ is the raw coincidence count, $\eta_{i} \in [0,1]$ is the joint pathway efficiency for coincidence $i$, and $\eta_{min}$ is the least of these joint pathway efficiencies. Since $\eta_{min}/\eta_{i}\leq 1$, normalization typically increases the uncertainty in the entanglement estimate.
We sum all coincidences that belong to the same coincidence type, $ss, sd, ds,$ or $dd$,
\begin{align}
k_{sd}&=c_{h_2h_3}'+c_{v_2v_3}' \\
k_{ss}&=c_{h_2h_2}'+c_{v_2v_2}'\label{countreduce1} \\
k_{ds}&=c_{h_2v_2}'+c_{h_3v_3}' \\
k_{dd}&=c_{h_2v_3}'+c_{h_3v_2}'\textrm{,}
\end{align}
to obtain coincidence type totals $\kappa$=$\left\{k_{sd},k_{ss},k_{ds},k_{dd}\right\}$. 
\par
The last technicality is that our experimental prototype monitors statistics in only one arm of the BSA, which reduces the number of observed same-port same-polarization coincidences by a factor of a half relative to the expected value. Comparing Eq. (\ref{countreduce0}) and (\ref{countreduce1}), we see that monitoring only one port for same-port same-polarization coincidences results in the absence of corrected counts $c_{h_3h_3}$ and $c_{v_3v_3}$ from Eq. (\ref{countreduce1}). Additionally, we use non-photon number resolving detectors, which reduces the recorded number of events by an additional factor of one half for these same-port same-polarization events. The joint efficiencies given above are determined relative to the actual number of photons in a given pathway. Thus, the joint efficiencies given above do not account for the losses relevant to same-port same polarization coincidence events. In principle, we could determine joint efficiencies that would account for these events, but it would greatly increase the uncertainty in our parameter. Instead, we maximize the certainty of our parameter estimation by including these known losses directly into the model. This is done by averaging over all possible values of the true number of same-port, same-polarization events $n_{ss}$. The resulting probability distribution for $\mathcal{P}$ given normalized coincidence type totals $\kappa$ is
\begin{equation}
P(\mathcal{P}|\kappa)=\frac{p_{sd}^{k_{sd}}p_{ds}^{k_{ds}}p_{dd}^{k_{dd}}\sum_{n_{ss}=k_{ss}}^{\infty}\binom{n_{ss}}{k_{ss}}\left(\frac{p_{ss}}{4}\right)^{n_{ss}}}{P(\kappa)}
\label{PD}
\end{equation}
$P(\kappa)$ is the numerator integrated over all possible values of coincident probabilities, $p_{sd},p_{ss},p_{ds},$ and $p_{dd}$.

Given the set of normalized coincidences totals $\kappa$, we estimate $\mathcal{E}$ as the mean
\begin{equation}
\mathcal{E}_\kappa=\int{P(\mathcal{P}|\kappa)(p_{dd}-p_{ds}) d\mathcal{P}}\\
\end{equation}
which yields
\begin{equation}
\mathcal{E}_\kappa=\frac{(k_{dd}-k_{ds})_2\tilde{F}_1(1\!+\!k_{ss},1\!+\!k_{ss};5\!+\!n;\frac{1}{4})}{_2\tilde{F}_1(1\!+\!k_{ss},1\!+\!k_{ss};4\!+\!n;\frac{1}{4})}
\end{equation}
where $n=k_{sd}\!+\!k_{ss}\!+\!k_{ds}\!+\!k_{dd}$ and $_2\tilde{F}_1(a;b;c)$ is the regularized hypergeometric function. We similarly calculate $\left(\mathcal{E}^2\right)_\kappa$ to derive the variance $\sigma_{\kappa}^{2}=\left(\mathcal{E}^2\right)_\kappa-\left(\mathcal{E}_\kappa\right)^2$
in the estimate.

\subsection{Tamper Detection}
Under normal operation, the quantum seal transmits polarization-entangled states through the active and reference fibers and the BSA measurements yield an estimate for the entanglement parameter $\mathcal{E}$. When the estimate $\mathcal{E}_{\kappa}$ exceeds the bound of 1/2, then the seal confirms that the received photon pairs are entangled. This use of quantum integrity verification certifies that the seal is unmolested. Moreover, the presence of entanglement is doubly indicative, as it confirms that the photon injected into the active fiber is both the same photon retrieved from the active fiber and that the path length difference between the reference and active fiber links is much smaller than the single-photon coherence length.
\par
By contrast, quantum integrity verification fails when the estimate $\mathcal{E}_{\kappa}$ lies below the threshold value of 1/2. This occurs in presence of tampering due to either a temporal shift in the photon time-of-arrival or a loss of entanglement between the photons from the intercept-resend attack. However, verification may also fail because of technical noise during transmission and measurement, e.g., decoherence of the entangled state. Therefore it is necessary to quantify the probability to accurately detect tampering as well as the rate at which detection fails.
\par
We formalize the tamper detection problem as a binary decision in which the estimate of $\mathcal{E}$ decides between two possible hypothesis \cite{vanTrees2001}. Under the positive hypothesis $H_1$, tampering is indicated when the entanglement estimate $\mathcal{E}_{k}$ is below a detection threshold defined as $\epsilon$. Under the null hypothesis $H_0$, the transmitted entanglement is preserved and the estimate exceeds the threshold $\epsilon$. We gain greater confidence in our decision by choosing $\epsilon > 1/2$ but at the expense of a higher probability of false alarm rate, i.e., classifying entangled states as being inauthentic.
\par
We characterize normal operation of the seal by an average entanglement parameter $\mathcal{E}_1$ and its corresponding variance $\sigma^2$. The nominal value $\mathcal{E}_1$ characterizes the amount of entanglement expected from the system in the absence of tampering. This value also characterizes the quality of the transmission and includes effects due to environmental decoherence. We also define the expectation value in the presence of tampering as $|\mathcal{E}_0|\leq 1/2$, which follows from the known the theoretical limitations of the entanglement parameter, and we will assume that the variance is the same as under normal operation, $\sigma_\kappa$. Given an estimate $\mathcal{E}_{\kappa}$, our task is to decide whether tampering has occurred,
\begin{equation}
\label{hypo0}
H_0 : \mathcal{E}_{\kappa} = \mathcal{E}_0 + e \leq \epsilon,
\end{equation}
or not,
\begin{equation}
\label{hypo1}
H_1 : \mathcal{E}_{\kappa} = \mathcal{E}_1 + e > \epsilon
\end{equation}
In both hypotheses, $e$ represents a zero-mean Gaussian random variable with average variance $\sigma_\kappa^2$, while the detection threshold $\epsilon$ will depend on the implementation. For example, the threshold value should be chosen to limit the number of false alarms as well as to limit the possibility of successful spoofing. The probability distribution for $\mathcal{E}_\kappa$ under each hypothesis is closely approximated by the Gaussian
\begin{equation}
P(\mathcal{E}_\kappa|\mathcal{E}_i,\sigma_\kappa)\approx\frac{1}{\sqrt{2\pi\sigma_\kappa^2}}\;\exp\!\left[-\;\frac{(\mathcal{E}_\kappa-\mathcal{E}_i)^2}{2\sigma_\kappa^2}\right]
\end{equation}
where $i\in\{0,1\}$.
\par
\begin{figure}[ht]
\begin{center}
\includegraphics[scale=0.6]{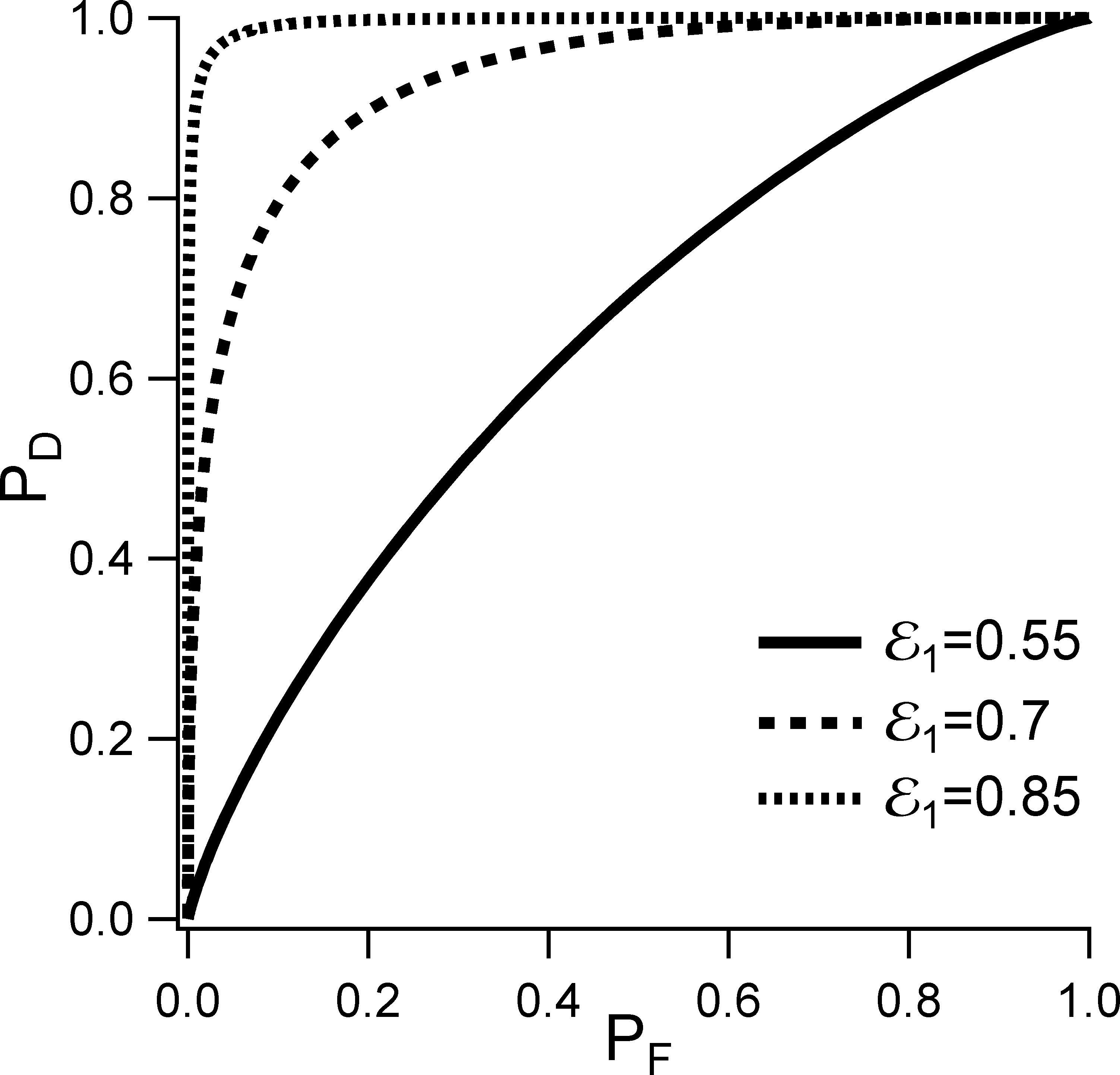}
\caption{The receiver operating characteristic (ROC) curves display a parametric plot of the probability of detection $P_D$ versus the false alarm rate rate $P_{FAR}$. Assuming that tampering yields $\mathcal{E}_0=1/2$ with uncertainty $\sigma=0.1$, we plot the behavior for baseline entanglement values of $\mathcal{E}_1 = 0.55,0.7,$ and $0.85$. Our experimental prototype exhibits much smaller uncertainty and higher baseline entanglement.}
\label{rocfig}
\end{center}
\end{figure}

The discrimination of two constant signals in the presence of Gaussian noise is a well known problem in binary detection theory \cite{vanTrees2001}. The probability to detect tampering is given as
\begin{equation}
\label{eq:pdet}
P_{D}= \frac{1}{2}\textrm{erfc}\left(\frac{\mathcal{E}_0-\epsilon}{\sqrt{2}\sigma_\kappa}\right)\textrm{.}
\end{equation}
where
\begin{equation}
\label{erfc}
\mathrm{erfc}(x) = \frac{2}{\sqrt{\pi}} \int_{x}^{\infty}{e^{-z^2} dz}
\end{equation}
is the complimentary error function. The corresponding probability for spoofing is
\begin{equation}
P_{S}=1-P_{D}
\end{equation}
while the probability of a false alarm, in which an authentic signal is misidentified as spoofed, is
\begin{equation}
P_{FAR}=\frac{1}{2} \textrm{erfc}\left(\frac{\mathcal{E}_1-\epsilon}{\sqrt{2}\sigma_\kappa}\right)
\end{equation}
The probability of detection and false alarm characterize operation of the seal for a given threshold $\epsilon$, and the parametric dependence of $P_D$ and $P_{FAR}$ are presented by the receiver operating characteristic curve in Fig.~(\ref{rocfig}). Therefore, operation of the seal can be tuned by selecting a detection threshold $\epsilon$ that provides a desired probability of detection or false alarm rate.
\par
It is apparent from Fig.~\ref{rocfig} that tampering can be detected with very high probability when the normal operating value $\mathcal{E}_1$ is close to the theoretical maximum of $\pm 1$. This is due to the sharp transition separating authentic and spoofed signal as represented by the bounds in Eq.~(\ref{bound}), which may be violated by many orders of standard deviation when transmitting the  $\Psi^{\pm}$ entangled state. Note that it is also possible that during the time $T$ required to estimate $\mathcal{E}$ that an intruder may inject an unentangled state for some duration $t<T$. The presence of the injected state would prepare a mixture of actual and spoofed states with the entanglement estimated as
\begin{equation}
\mathcal{E}(t,T)=\frac{t}{T}\mathcal{E}_0+\frac{(T-t)}{T}\mathcal{E}_1
\end{equation}
Intrusions would be indicated when $\mathcal{E}(t,T) < \epsilon$, i.e., for
\begin{equation}
t>\frac{\left(\mathcal{E}_1-\epsilon\right)}{\left(\mathcal{E}_1-\mathcal{E}_0 \right)} T \nonumber\textrm{.}
\end{equation}
Since $\mathcal{E}_{0}$ may be at most $1/2$ and $\mathcal{E}_1$ is a constant of the system, the duration for this short-time spoofing decreases linearly with the threshold $\epsilon$. Moreover, the probability of detection increases nonlinearly with $\epsilon$, as seen in Eq.~(\ref{erfc}), and short-time attacks can be detected with high probability at a relatively low false alarm rate using only moderate increases in detection threshold. Short-time attacks may also be mitigated by decreasing the duration $T$ used for estimating the entanglement. 
\par
While entanglement excludes spoofing the measurement apparatus with a cloned state, a savvy adversary may attempt to redirect the active photon away from the tamper seal in order to access the surveyed area. The estimated entanglement in this scenario will behave normally when the optical path length of the redirected photon is exactly matched. We address this vulnerability in our implementation by making the entanglement estimate sensitive to sub-millimeter disturbances in the expected path length. Such tolerances can be adjusted to fit application-specific requirements, for example, by broadening the spectral bandwidth of the single-photon states emitted from the source. Temporal shaping will retain the spectral entanglement required for tamper detection and converge on interferometric fringe stability in the infinite bandwidth limit. This offers greater system stability as well as increased sensitivity to the redirection attack.
\par
Depending on the length of the optical fiber used and the location of deployment, variations in the ambient temperature may alter the active and reference fiber path lengths. Fibers exposed to different ambient conditions may experience a change in path length $\Delta$L (m) that is proportional to the initial fiber length L$_0$ (m) and temperature change $\Delta$T ($\degree$C),
\begin{equation}\Delta\textrm{L} =  \frac{\alpha\textrm{L}_0}{\textrm{m}\cdot\degree\!\textrm{C}} \Delta\textrm{T}\end{equation}
with thermal expansion coefficient $\alpha=10^{-6}$ being a high estimate \cite{bachmann1988thermal}. Consider the example of  L$_0$=1 km for each of the active and reference fibers and a temperature differential $\Delta$T=$\pm$ 10$\degree$C. This produces a round-trip path length change of $\Delta$L=$\pm$ 20 mm. This change is large enough to distinguish the active and reference photons temporally while also small enough to be compensated by simple adjustments to the reference path length. For example, a free-space coupler mounted on a translatable stage can be used to regulate the path length alongside environmental monitoring. 
\par
The expected changes in path length due to time of day and weather patterns are relatively slow and very distinct from the sudden, unexpected changes in path length that result from tampering. The quantum seal is sensitive to effects on the time scale $T$, and this includes sudden changes in environmental conditions such as  pressure on the fiber or a rapid rise in temperature. The seal does not discriminate between intentional and unintentional acts that modify decoherence in the transmission channel. However, our theoretical account of stochastic fluctuations on the estimate demonstrate that the probability of detection remains high even for moderate value of $\sigma_{\kappa}$, cf. Eq.~(\ref{eq:pdet}).
\par
We also examine the attack in which an intruder replaces the active link with an exact replica. This replacement attack may be technically challenging, but it is nonetheless physically possible. Implementing the attack would require first characterizing the active fiber link, then severing the original transmission channel and joining the replica fiber to both the transmitter and receiver, and finally coupling the photon into the replica link. These steps are easily detectable as they modify the physical properties of the fiber and break the continuity of transmission. Alternatively, the seal system could be depowered so that optical continuity was broken at the source. A blackout event would again indicate tampering and when power was returned to the system, a system calibration procedure that included checking fiber installation would reveal evidence of the replacement attack.
\section{Experimental Implementation}\label{experimentsec}
\begin{figure}[b]
\begin{center}
\includegraphics[scale=0.38]{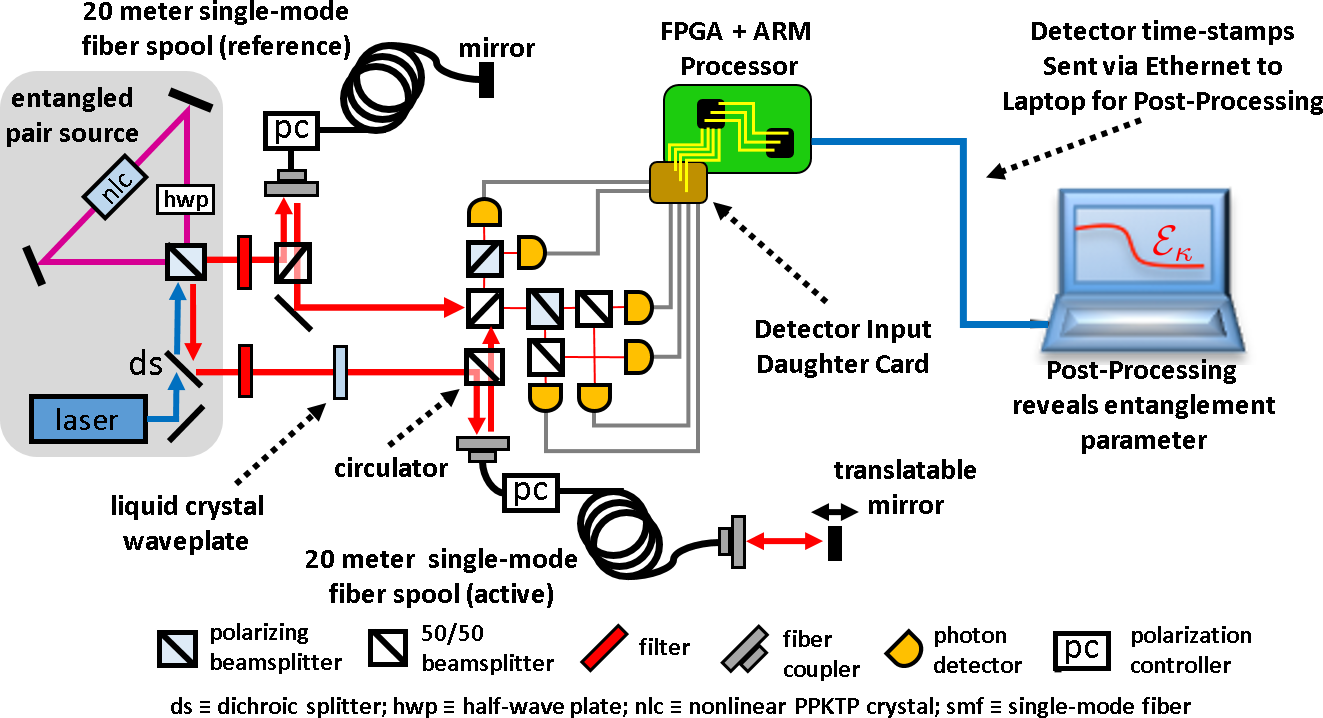}
\caption{The experimental schematic for our prototype tamper-indicating quantum seal showing its four major components: (a) a polarization-entangled photon source, (b) 20 meter single-mode fiber-optic channels for the reference and active links, (c) polarization-entanglement verification based on Bell-state analyzer, and (d) timestamping electronics integrated with a processor to transmit observed coincidences to a host computer for parameter estimation.}
\label{prototype}
\end{center}
\end{figure}

In this section, we detail the experimental implementation of the quantum seal modeled in Sec.~\ref{theorysec} and Fig.~\ref{figschematic}. A schematic of the experimental layout is shown in Fig.~\ref{prototype}, in which the entangled light source is based on cw-pumping of type-II spontaneous parametric down-conversion (SPDC) in a nonlinear optical PPKTP (periodically polled potassium titanyl phosphate) crystal within a Sagnac loop. This configuration generates a polarization-entangled pair state $\Psi^+$ \cite{Kim2006}. 
The narrowband pump laser operates at 405 nm and a power of approximately 1 mW. At this power, the source generates approximately 1.2 million photon pairs per second. The nonlinear PPKTP crystal is 30 mm in length and phase-matched to produce near-degenerate energy collinear photon pairs with orthogonal polarizations. These photons have a central wavelength centered an 810 nm with a 0.5 nm bandwidth.

The entangled light source transmits each photon of the entangled pair to separate polarization phase-maintaining circulators. For convenience, we use lossy, but easily configured, symmetric beamsplitters as circulators. As opposed to more elaborate setups, this reduces the overall photon pair collection efficiency by $90\%$ and represents the dominant loss mechanism. Whereas this setup is sufficient for our proof-of-principle demonstration, future tamper seals would benefit from using more efficient dual polarization phase preserving optical circulators. Following the circulators, one photon is output to the active fiber, while the other is routed to the reference fiber. We use 20-m single-mode fibers for both the active and reference links with polarization controllers to correct for polarization rotation that occurs during the fiber round-trip. Polarization entanglement would be reduced or lost completely without this correction. We include an optical delay stage at the terminal of the active fiber in order to implement redirection attacks.

Once the active and reference photons have completed the round trip through the respective fibers, they are routed to a Bell-state analyzer consisting of symmetric beamsplitters, polarization beamsplitters, and single-photon detectors. Our BSA is modeled after the design shown in Fig.~\ref{BSA}, but we do not monitor the full set of possible coincidences. This is due to the symmetry expected for the detection probabilities given by Eqs.~(\ref{prob0})-(\ref{prob3}). This comes at the expense of a reduced data set size and correspondingly greater uncertainty in the parameter estimate, but reduces the number of single-photon detectors needed. We use Perkin-Elmer (now Excelitas) SPCM devices which have an efficiency of 0.4 at photon wavelength 810 nm. Due to pathway efficiencies ranging from $5-6\times10^{-3}$ we observe single-photon counts per second (cps) in the range of $2-6 \times 10^3$ cps and average coincidence rates in the range of 0-15 cps.

We measure photon time-of-arrival by monitoring the output signal from the detector array. Photon detection triggers a detector to output a 25 ns TTL pulse that is then timestamped and logged according to detector identifier. We implement timestamping by sending the TTL pulses to an FPGA (field programmable gate array) configured with a custom daughter card that connects up to eight coaxial cables to input pins. We have reported previously on the use of FPGA's for performing single-photon detection, in which we timestamp each TTL pulse against the FPGA clock and store the resulting data alongside the input channel identifier \cite{pooser2012fpga}. In the current implementation, we use a Zedboard system-on-a-chip composed from an FPGA and ARM processor.\footnote{http://www.zedboard.org} We operate the FPGA using a 10 ns clock cycle for the detection system. Stored timestamp data is then aggregated and packetized by custom software running on the ARM processor within the Xillinux operating system.\footnote{http://www.xillybus.com} These packets are then transmitted over Ethernet using the UDP protocol to a host computer where the entanglement parameter is estimated \cite{Humble2014}. 

\begin{figure}[t]
\centering
\vspace{10pt}
\includegraphics[scale=0.38]{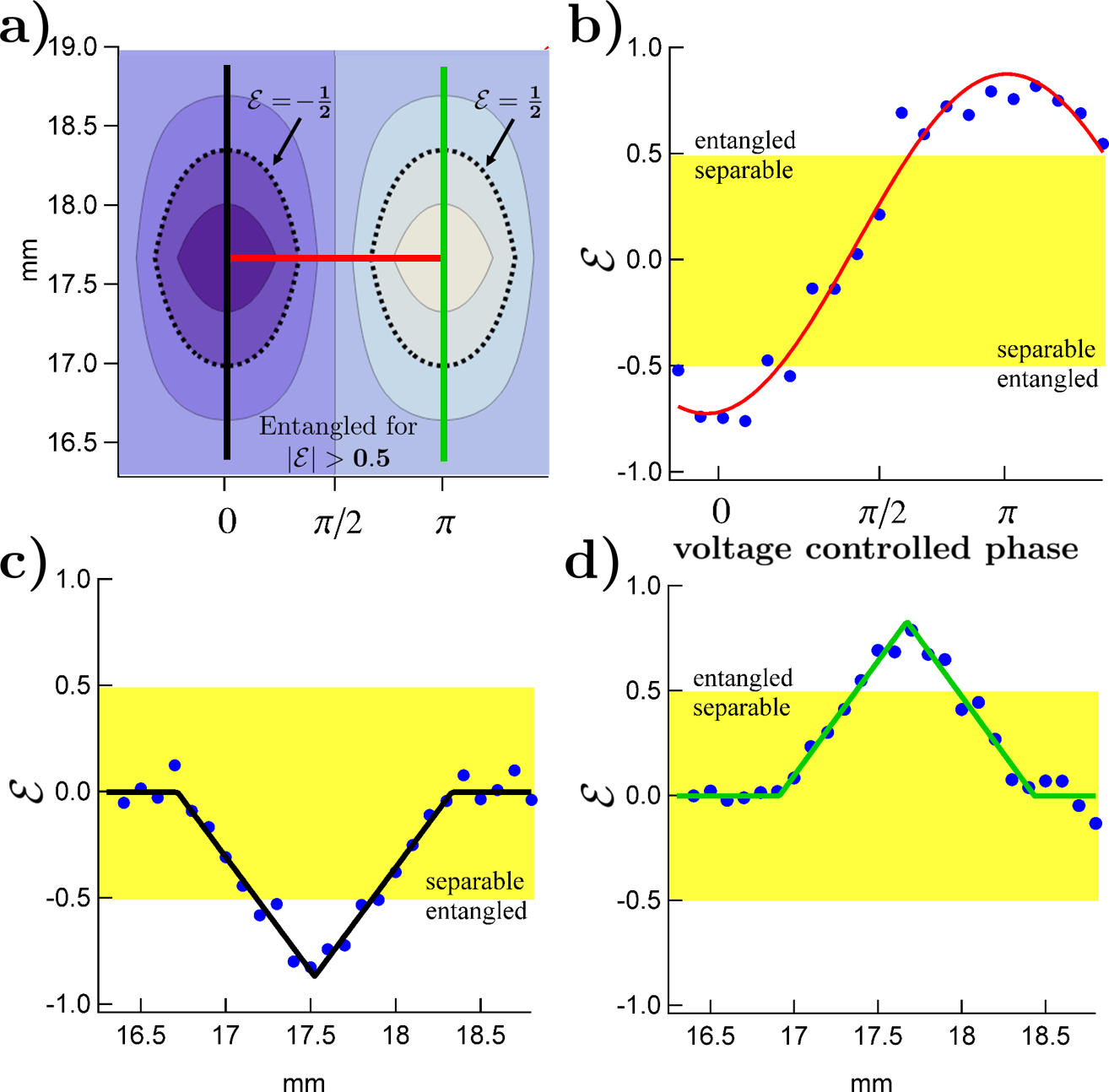}
\caption{a) This contour plots the theoretical value of parameter $\mathcal{E}$ versus phase and path length delay (mm). Red, black, and green dotted lines indicate the range over which the experimental data in plots b), c), and d) are taken. b) Experimental values for $\mathcal{E}$ as the phase of the polarization entangled state is modulated with a liquid crystal waveplate. At $0$ and $\pi$ phase the $\Psi^-$ and $\Psi^+$ states are obtained, respectively. c) Experimental values for $\mathcal{E}$ in the state $\Psi^-$ as the active photon's path length is increased length. d) Experimental values for $\mathcal{E}$ in state $\Psi^+$ as the active photon's path length is increased length.}
\label{tile}
\end{figure}

We experimentally detect tampering by estimating $\mathcal{E}$ using the prototype quantum seal. The results are shown in Fig.~\ref{tile}, where the dependence of $\mathcal{E}_{\kappa}$ is plotted against tuning of the entanglement in the input state and the relative difference between the active and reference paths. These plots correspond to modifying the delay and phase parameters of Eq.~(\ref{correlation3}). Assuming the source generates the $\Psi^+$ Bell state, the entanglement parameter is
\begin{equation}
\mathcal{E}={\textrm{\Large$\wedge$}}(2t_d/\tau_c)\mathrm{Cos}\left(\phi+\pi\right)\textrm{.}
\end{equation}
which clearly depends on the optical delay $t_d$ and the relative phase $\phi$. Settings of $\phi$=$0$ and $\phi$=$\pi$ correspond to the Bell states $\Psi^-$ and $\Psi^+$ , respectively. By adjusting the phase $\phi$, we tune the state that was returned to the detector from being authentic to inauthentic. This was accomplished experimentally by modulating, changing the applied voltage, a liquid crystal waveplate in the active photon path as seen in Fig. \ref{prototype}. The temporal delay $t_d$ was adjusted by translating the reflection mirror at the end of the active fiber.
\par
The experimentally estimated entanglement parameter is shown in Fig.~\ref{tile}, where a maximum value of $|\mathcal{E}_\kappa|\approx 0.8$ is obtained. This statistically significant estimate is well above the separable bound of $0.5$ and a clear signature of entanglement as nominal behavior. Figure~\ref{tile}(b) shows how the estimate varies as the phase of prepared state is varied and the detected state is rotated from $\Psi^{+}$ to $\Psi^{-}$. In addition, Figs.~\ref{tile}(c) and (d) show how sub-millimeter path length delays lead to decreases in the estimated entanglement.
\par
For the system demonstrated here, each entanglement estimate $\mathcal{E}_{\kappa}$ is calculated from coincidences collected during a 10 second sampling window interval. For this sampling interval, we observe combined raw coincident counts of $c_{i}\in\{ 0,400\}$  with corresponding standard deviation $\sigma_{\kappa} = 0.03-0.04$. Using these experimentally measured parameters, the detection theory presented in Section \ref{theorysec} yields greater than 0.9999 probability of detecting inauthentic signals with a false alarm chance of $10^{-9}$ when using 10 second sampling interval. Longer sampling windows lead to even lower false alarm rates.
\par
Several factors contribute to reduction in the entanglement estimate from a maximal value of 1. Foremost is the imperfect correction offered by the polarization controllers, which are subject to drift during acquisition time. The use of Faraday rotators at the fiber terminals can provide better correction for this type of system noise. Interfaces between the optical fibers and free-space optics also degrade the entanglement, with unwanted reflections causing photon pairs to become temporally distinguishable and polarization uncorrected. These resulted in unwanted coincidence events that did not participate in HOM interference. Future improvements to the seal design should make use of more stable Faraday mirrors to overcome the need to correct the polarization, and angled optical fiber connectors to eliminate the unwanted reflections.

\section{Conclusions}\label{conclusionsec}
We have reported on the design, operation, and implementation of a tamper-indicating quantum seal. Our approach is based on transmitting polarization-entangled photon pair states over optical fibers and monitoring the received entanglement in near real-time. We estimate the received entanglement from measured coincidences in a Bell-state analyzer. We show that the entanglement parameter $\mathcal{E}$ is sufficient to discriminate between the entangled states transmitted by the seal and those states that have been modified by an intruder.
\par
We have also presented a detailed theoretical model accounting for the seal operation including its physics and an analysis of its performance in terms of binary detection theory. We have also provided a detailed experimental implementation of the major subsystems, including an entangled photon pair source based on cw-pumped SPDC and a Bell-state analyzer based on two-photon interferometric time-of-arrival measurements. We have integrated these subsystems with an entanglement monitoring system, and we have reported the performance of this seal implementation in terms of the generated entanglement, acquisition time, state phase, and temporal delay as they relate to entanglement detection sensitivity. Based on these results, we conclude that the seal detects spoofed states with a probability $(P_D > 0.9999)$ at a very low false alarm rate $(P_{F} \ll 10^{-9})$ for a 10 second time interval. In conclusion, the tamper-indicating quantum seal presented here offers unprecedented surety in the detection of intrusion against fiber optical seal systems. Applications in containment and surveillance technologies, as well as telecommunications security and fiber-based sensors, are likely to benefit from the adaption of these ideas. 

\section*{Acknowledgments}
This work was supported by the Defense Threat Reduction Agency. This manuscript has been authored by UT-Battelle, LLC, under Contract No. DE-AC05-00OR22725 with the U.S. Department of Energy. The United States Government retains and the publisher, by accepting the article for publication, acknowledges that the United States Government retains a non-exclusive, paid-up, irrevocable, world-wide license to publish or reproduce the published form of this manuscript, or allow others to do so, for United States Government purposes.

\bibliography{Quantum.Optical.Seals}
\end{document}